\documentclass[aip,pop,preprint,a4paper]{revtex4-1}

\usepackage{graphicx}
\usepackage{bm}
\usepackage{MnSymbol}
\usepackage{setspace}

\usepackage{xcolor}

\begin{document}

\title{Fast ion effects on the threshold conditions of ion temperature gradient mode and electron temperature gradient mode}

\author{M.K. Jung}
\affiliation{Department of Nuclear Engineering, Seoul National University, Seoul 08826, South Korea}
\author{T.S. Hahm}
\email[]{tshahm@snu.ac.kr}
\affiliation{Nuclear Research Institute for Future Technology and Policy, Seoul National University, Seoul 08826, South Korea}
\author{Y.-S. Na}
\affiliation{Department of Nuclear Engineering, Seoul National University, Seoul 08826, South Korea}
\author{E.S. Yoon}
\affiliation{Department of Nuclear Engineering, Ulsan National Institute of Science and Technology, Ulsan 44919, South Korea}

\date{\today}

\onehalfspacing
\begin{abstract}
We investigate the fast ion effects on the threshold conditions of ion temperature gradient (ITG) mode and electron temperature gradient (ETG) mode both analytically and numerically using gyrokinetic equation. The onset condition for ITG mode shows a strong and monotonic favorable dependence on the fraction of fast ions, and mostly favorable but non-monotonic dependence on the fast ions' normalized temperature $T_f/T_i$ ($T_f$ is the effective temperature of fast ions, $T_i$ is the temperature of thermal ions). Overall favorable parametric trends are consistent with those for the linear growth rate reported in previous papers, as they are largely determined by kinetic wave-particle resonance effects. While general analytic expressions for the critical normalized thermal ion temperature gradient scale length $(R/L_{T_i})_c$ are quite complicated, an explicit compact expression $\left(\frac{R}{L_{T_i}}\right)_c=\left(\frac{4}{3}+\frac{3}{2}\sqrt{\frac{\pi}{2}}\frac{|\hat{s}|}{q}\right)\left(1+\frac{T_i}{Z_i(1-f_h)T_e}\right)$ has been derived for the mode with its perpendicular scale larger than thermal ion gyroradius, but much smaller than the fast ion gyroradius so that finite Larmor radius effects are manifested in opposite asymptotic limits depending on ion species when $T_f\gg T_i$, and weak density gradient. Here, $q$ is safety factor, $\hat{s}$ is magnetic shear, $Z_i$ is thermal ions' charge, and $f_h$ is fast ion charge density fraction. In this limit, only the fast-ion-induced thermal ion dilution effects persist as fast ion density response becomes unmagnetized and negligible. On the other hand, the fast ion effects on ETG-threshold are found to be unfavorable.
\end{abstract}
\doublespacing
\maketitle

\section{Introduction}\label{sec:introduction}

Fast ions are generated by auxiliary heating such as NBI (neutral beam injection) and ICRH (ion cyclotron resonance heating) in present day magnetically confined plasmas, and by fusion reactions in future devices. While fast ions can excite various instabilities and sometimes degrade plasma confinement, accumulating evidence over the past decade indicates that fast ions can also significantly enhance confinement through turbulence reduction. For example, internal transport barriers (ITBs) have been formed and sustained for up to 50 s in the ion channel on KSTAR in the so-called FIRE mode, where NBI is used as the primary heating scheme. \cite{HanN22,HanPoP24,NaNF26} Confinement enhancement in the presence of fast ions produced by NBI or ICRH has been reported across several tokamak experiments which include ASDEX Upgrade, \cite{TardiniNF07,DiSienaPRL21} DIII-D, \cite{LiuPRL22,BrochardPRL24,DuPRL25} EAST, \cite{LiuNF20} HL-2A, \cite{LinNF23} JET, \cite{RomanelliPPCF10,ManticaPRL11,GarciaNF13,MazziNP22} and KSTAR. \cite{NaNF20,LeeNF23,KimNF23} Understanding physical mechanisms behind these findings is currently a hot topic, and a very broad subject. Progress in the past decade has been recently reviewed in Ref.~\onlinecite{NaNRP25}.

In this paper, we focus on one of the simplest, but relevant considerations addressing the fast ion effects on turbulence, i.e. the influence of fast ions on the onset condition of ITG (ion temperature gradient) instability. ITG instability is a primary candidate responsible for most of anomalous ion thermal transport. Since it can cause significant transport once linearly excited, its linear threshold condition is a quantity of relevance and interest in confinement research. In fact, it has been one of the first steps in validating microinstability theory predictions against experimental observations for more than three decades. \cite{ScottPRL90,StambaughPFB90,KimNF18,KimNF20,DingPoP21,KnolkerPPCF21}

Naturally, there have been numerous gyrokinetic simulation studies investigating the favorable role of fast ions on ITG stability. \cite{TardiniNF07,DiSienaNF18,DiSienaPoP19,HanN22,WangPST22,KimNF23,IshizawaCP25} Most of those focused on behavior of the linear growth rates and induced transport. To our knowledge, there has been little work addressing the effects of fast ions on ITG onset condition. Therefore, it would be useful to investigate the influence of fast ions on ITG threshold covering a wide range of parameter space (in particular, the fraction of fast ion population $f_h$ and the effective temperature, quantified by the mean kinetic energy $T_f$ of fast ions) relevant to present day major experiments and future devices such as ITER.

In this paper, we derive the ITG linear threshold condition for instability, $R/L_{T_i}>(R/L_{T_i})_c$ analytically in the presence of considerable fast ion population characterized by the effective temperature $T_f$ exceeding several $T_i$. We also perform linear gyrokinetic simulations using \texttt{GENE} \cite{JenkoPoP00} code, and compare to ensure validity of the analytic results. We consider two ion species, i.e. thermal and fast ions, described by linear electrostatic gyrokinetics and adiabatic electrons in either concentric circular toroidal geometry or in sheared slab geometry. We follow the methods employed for previous studies in the absence of fast ions. \cite{HahmPFB89,RomanelliPFB89,YoonNF10}

The principal results of this paper are as follows.
\begin{enumerate}
    \item[(i)] $(R/L_{T_i})_c$ increases with $f_h$ strongly and monotonically for other parameters typical of present day major experiments and ITER kept constant.
    \item[(ii)] $T_f/T_i$-dependence of $(R/L_{T_i})_c$ is more complicated. For $\eta_f\equiv\frac{L_{n_f}}{L_{T_f}}>\frac{2}{3}$ which is typical of ICRH case, $(R/L_{T_i})_c$ increases with $T_f/T_i$ for moderate values of $T_f/T_i$ at fixed $f_h$, then takes a maximum value for $T_f/T_i\sim10$. Then, it decreases mildly as $T_f/T_i$ increases further. For $\eta_f<\frac{2}{3}$ which is typical of NBI case, $(R/L_{T_i})_c$ increases with $T_f/T_i$ monotonically at constant $f_h$, approaching a constant value determined by the dilution effect only at very high $T_f/T_i$. This trend is similar to that of the linear growth rate $\gamma_{lin}$ reported in previous publications. \cite{DiSienaNF18,DiSienaPoP19}
    \item[(iii)] The threshold $(R/L_{T_i})_c$ increases with the normalized magnetic shear $\hat{s}/q$ as indicated from the result in the sheared slab limit. Here, $q$ is safety factor and $\hat{s}\equiv r/q\cdot dq/dr$ is magnetic shear.
    \item[(iv)] For $k_\perp\rho_f>1$, fast ion response becomes unmagnetized and only the dilution effects persist. A compact analytic formula $\left(\frac{R}{L_{T_i}}\right)_c=\left(\frac{4}{3}+\frac{3}{2}\sqrt{\frac{\pi}{2}}\frac{|\hat{s}|}{q}\right)\left(1+\frac{T_i}{Z_i(1-f_h)T_e}\right)$ has been obtained by combining result of the flute-like mode in the sheared slab and that of the strongly ballooning mode. This result can be relevant in the presence of fusion-product fast ions expected in ITER and future devices, and for some present day ICRH experiments. \cite{MazziNP22}
    \item[(v)] In addition, fast ion effects on ETG (electron temperature gradient) threshold condition has been investigated both analytically and numerically. An analytic expression $\left(\frac{R}{L_{T_e}}\right)_c=\left(\frac{4}{3}+\frac{3}{2}\sqrt{\frac{\pi}{2}}\frac{|\hat{s}|}{q}\right)\left(1+Z_i(1-f_h)\frac{T_e}{T_i}+Z_ff_h\frac{T_e}{T_f}\right)$ summarizes the parametric dependences for weak density gradients. Gyrokinetic simulation leads to minor modifications of the first factor to $\left(1.3+1.6\frac{|\hat{s}|}{q}\right)$. In contrast to the ITG case, the results indicate an unfavorable influence.
\end{enumerate}

The remainder of this paper is organized as follows. The theoretical model leading to governing equations for the threshold calculation is introduced in Sec.~\ref{sec:model}. In Sec.~\ref{sec:slab}, flute-like acoustic branch of ITG mode is considered in sheared slab geometry. In Sec.~\ref{sec:toroidal}, strongly ballooning ITG mode in toroidal geometry is considered. Numerical simulation results are presented in Sec.~\ref{sec:simulation}, and conclusions are drawn in Sec.~\ref{sec:conclusion}. Fast ion effects on ETG threshold condition is presented in the Appendix.

\section{Theoretical Model}\label{sec:model}

Our model involves three species: electrons, main (thermal) ions, and fast ions. We will be denoting the species with a subscript e, i, and f.

We start from the conservative modern nonlinear gyrokinetic Vlasov equation
\begin{equation}\label{eqn1}
    \frac{\partial f}{\partial t}+\frac{d\bm{R}}{dt}\cdot\frac{\partial f}{\partial\bm{R}}+\frac{dv_\parallel}{dt}\frac{\partial f}{\partial v_\parallel}=0,
\end{equation}
where $f$ is the gyrocenter distribution function. \cite{HahmPF88} After splitting $f$ into $f_0+\delta f$, where $f_0$ and $\delta f$ are the equilibrium and perturbed distribution function, respectively, linearization of Eq.~(\ref{eqn1}) leads to
\begin{equation}\label{eqn2}
    -\left[\frac{\partial}{\partial t}+\left(\frac{d\bm{R}}{dt}\right)_0\cdot\frac{\partial}{\partial\bm{R}}+\left(\frac{dv_\parallel}{dt}\right)_0\frac{\partial}{\partial v_\parallel}\right]\delta f=\left[\left(\frac{d\bm{R}}{dt}\right)_1\cdot\frac{\partial}{\partial\bm{R}}+\left(\frac{dv_\parallel}{dt}\right)_1\frac{\partial}{\partial v_\parallel}\right]f_0,
\end{equation}
where the subscripts indicate ordering. The explicit expressions for each term are
\begin{eqnarray}
    \left(\frac{d\bm{R}}{dt}\right)_0=v_\parallel\frac{\bm{B^{*}}}{B^*}+\frac{cm\mu}{ZeB^*}\bm{b}\times\nabla B \label{eqn3},\\
    \left(\frac{dv_\parallel}{dt}\right)_0=-\frac{\bm{B^{*}}}{B^*}\cdot\mu\nabla B, \\
    \left(\frac{d\bm{R}}{dt}\right)_1=\frac{1}{B^*}c\bm{b}\times\nabla\llangle\delta\phi\rrangle, \\
    \left(\frac{dv_\parallel}{dt}\right)_1=-\frac{\bm{B^{*}}}{mB^*}\cdot Ze\nabla\llangle\delta\phi\rrangle, \label{eqn6}
\end{eqnarray}
where $\bm{B^{*}}\equiv\bm{B}+(cm/Ze)v_\parallel\nabla\times \bm{b}$, $B^*\equiv\bm{B^{*}}\cdot\bm{b}$, $\mu\equiv v_\perp^2/2B$, $v_\parallel$ and $v_\perp$ are velocity parallel and perpendicular to the magnetic field, $m$ and $Z$ are the mass and charge of the particle, $c$ is speed of light, and $\llangle\delta\phi\rrangle$ is the gyro-averaged perturbed electric potential. For low-$\beta$ equilibrium, $\nabla\times\bm{b}\approx\bm{b}\times\nabla\ln B$, and thus $B^*=B+(cm/Ze)v_\parallel\bm{b}\times\nabla\ln B\cdot\bm{b}\approx B$. Note that this does not imply that $\bm{B^{*}}=\bm{B}$. In the present model, we adopt the electrostatic approximation and therefore neglect electromagnetic perturbations, which is appropriate for low-$\beta$ plasmas.

By plugging Eqs.~(\ref{eqn3})-(\ref{eqn6}) into Eq.~(\ref{eqn2}), the expression for the perturbed distribution function of gyro-center can be obtained:
\begin{equation}\label{eqn7}
   \delta f_k=\frac{\frac{c}{B}J_0\delta\phi_k\bm{b}\times\bm{k}\cdot\frac{\partial f_0}{\partial\bm{R}}-\left(\frac{Ze}{m}k_\parallel J_0\delta\phi_k+\frac{c}{B}v_\parallel J_0\delta\phi_k\nabla\times \bm{b}\cdot\bm{k}\right)\frac{\partial f_0}{\partial v_\parallel}}{\omega_k-k_\parallel v_\parallel-\omega_{curv,k}-\omega_{\nabla B,k}},
\end{equation}
where $\bm{k}$ is the wave vector of perturbed field, $\omega_{curv,k}\equiv(cmv_\parallel^2/ZeB)\bm{b}\times(\bm{b}\cdot\nabla)\bm{b}\cdot\bm{k}$ and $\omega_{\nabla B,k}\equiv(cm\mu/ZeB)\bm{b}\times\nabla B\cdot\bm{k}$ are the curvature and $\nabla B$ drift frequencies, and $J_0=J_0(k_\perp\rho)$ is the zeroth order Bessel function, where $\rho$ is each particle's gyroradius.

Assuming that the equilibrium distribution function is Maxwellian with isotropic temperature (i.e. $T_\perp=T_\parallel=T$) and $v_T^2\equiv T/m$, 
\begin{equation*}
    f_0\equiv\frac{n_0}{(2\pi v_{T}^2)^{3/2}}\exp\left(-\frac{mv_\parallel^2/2+m\mu B}{T}\right),
\end{equation*}
we get
\begin{equation}\label{eqn8}
    \frac{\partial f_0}{\partial\bm{R}}=\left[\nabla\ln n_0+\left(\frac{v_\parallel^2/2+\mu B}{v_T^2}-\frac{3}{2}\right)\nabla \ln T-\frac{\mu B}{v_T^2}\nabla\ln B\right]f_0
\end{equation}
and
\begin{equation}\label{eqn9}
    \frac{\partial f_0}{\partial v_\parallel}=-\frac{v_\parallel}{v_T^2}f_0.
\end{equation}

By substituting Eqs.~(\ref{eqn8})-(\ref{eqn9}) to Eq.~(\ref{eqn7}), the non-adiabatic part of the perturbed gyrocenter distribution function $\delta h=\delta f+(ZeJ_0\delta\phi/T)f_0$ can now be written as
\begin{equation}\label{eqn10}
    \delta h_k=\frac{\omega_k-\omega_{*T,k}}{\omega_k-k_\parallel v_\parallel-\omega_{curv,k}-\omega_{\nabla B,k}}\frac{ZeJ_0\delta\phi_k}{T}f_0,
\end{equation}
where
\begin{equation}\label{eqn11}
    \omega_{*T,k}\equiv\frac{v_T^2}{\Omega}\bm{b}\times\nabla\ln n_0\cdot\bm{k}\left[1+\left(\frac{v_\parallel^2/2+\mu B}{v_T^2}-\frac{3}{2}\right)\eta\right].
\end{equation}
Here, $\Omega\equiv ZeB/cm$ is the cyclotron frequency, and $\eta\equiv\frac{\partial}{\partial r}\ln T_0/\frac{\partial}{\partial r} \ln n_0$. In this section, we are keeping both the drift frequencies $\omega_{curv}, \omega_{\nabla B}$ and parallel transit frequency $k_\parallel v_\parallel$. Further subsidiary frequency ordering will be made in Sec.~\ref{sec:slab} and~\ref{sec:toroidal}.
The perturbed electron density is assumed to be adiabatic, for $k_\parallel v_{T_e}\gg\omega$,
\begin{equation}\label{eqn12}
    \delta n_e=\frac{e\delta\phi}{T_e}n_{e0}.
\end{equation}
For the thermal ions and fast ions, after considering the polarization density and pull-back transformation from gyro-center to particle variables, \cite{HahmPF88} the perturbed particle densities consist of adiabatic and non-adiabatic part:
\begin{eqnarray}
    \delta n_i=\frac{e\delta\phi}{T_i}(1-f_h)n_{e0}-2\pi\int B^*\delta h_iJ_0d\mu dv_\parallel \label{eqn13},\\
    \delta n_f=\frac{e\delta\phi}{T_f}f_hn_{e0}-2\pi\int B^*\delta h_fJ_0d\mu dv_\parallel.\label{eqn14}
\end{eqnarray}

Then imposing the quasi-neutrality condition on Eqs.~(\ref{eqn12})-(\ref{eqn14}) gives the following relation
\begin{equation}\label{eqn15}
   \left[\frac{1}{\tau}+Z_i(1-f_h)+Z_ff_h\frac{T_i}{T_f}\right] \frac{n_{e0}e\delta\phi}{T_i}-2\pi Z_i\int B^*\delta h_iJ_0d\mu dv_\parallel-2\pi Z_f\int B^*\delta h_fJ_0d\mu dv_\parallel=0,
\end{equation}
where $\tau\equiv T_e/T_i$.
We will consider the weak electron density gradient limit which is justified for $L_{n_e}\gtrsim(RL_{T_i})^{1/2}$, unless specified otherwise.

\section{Flute-like acoustic branch in sheared slab geometry}\label{sec:slab}

We consider a flute-like acoustic branch of ITG mode \cite{CoppiPF67} in sheared slab geometry first. We begin with Eq.~(\ref{eqn10}) and consider a sheared slab geometry in which the equilibrium magnetic field is given by $\vec{B}_0=B_0(\hat{z}+\frac{x}{L_s}\hat{y})$, where $L_s\equiv qR/\hat{s}$. Therefore, $\omega_{curv},\omega_{\nabla B}$ are ignored in this section. Note that within this model negative magnetic shear case is just a mirror image of the positive magnetic shear case, so one should interpret $\hat{s}$ as $|\hat{s}|$. This remark applies to the results of Ref.~\onlinecite{HahmPFB89}. Then,
\begin{equation}\label{eqn16}
    \delta h_k=\frac{\omega_k-\omega_{*T,k}}{\omega_k-k_\parallel v_\parallel}\frac{ZeJ_0\delta\phi_k}{T}f_0.
\end{equation}
Now, the quasi-neutrality relation given in Eq.~(\ref{eqn15}) becomes
\begin{eqnarray}\label{eqn17}
    \left[\frac{1}{\tau}+Z_i(1-f_h)+Z_ff_h\frac{T_i}{T_f}-\frac{Z_i(1-f_h)}{\sqrt{2\pi}}\int_0^\infty{\frac{1}{v_{T_i}^3}\frac{\omega-\omega_{*T_i}}{\omega-k_\parallel v_\parallel}J_0^2e^{-v^2/2v_{T_i}^2}dv^3}\right]\delta\phi \nonumber\\
    -\left[\frac{Z_ff_h}{\sqrt{2\pi}}\frac{T_i}{T_f}\int_0^\infty{\frac{1}{v_{T_f}^3}\frac{\omega-\omega_{*T_f}}{\omega-k_\parallel v_\parallel}J_0^2e^{-v^2/2v_{T_f}^2}dv^3}\right]\delta\phi=0.
\end{eqnarray}
In the long wavelength limit of $k_\perp^2\rho_i^2\ll k_\perp^2\rho_f^2\ll1$, Eq.~(\ref{eqn17}) can be reduced to a differential equation in radial direction. Here, $\rho_i=\sqrt{m_iT_i}/Z_ieB$ and $\rho_f=\sqrt{m_fT_f}/Z_feB$ are gyroradii of thermal and fast ions at average kinetic energy, respectively. Furthermore, near ITG threshold, $|\omega|\ll|k_\parallel|v_{T_i}\ll|k_\parallel|v_{T_f}$ is satisfied, \cite{HahmPFB89} which simplifies the equation to
\begin{eqnarray}\label{eqn18}
    \left[\frac{1}{\tau}+Z_i(1-f_h)+Z_ff_h\frac{T_i}{T_f}+\frac{Z_i(1-f_h)}{2\sqrt{2\pi}}\int_{-\infty}^\infty{\int_0^\infty{\frac{\omega_{*T_i,k}/k_\parallel v_{T_i}}{\omega_k/k_\parallel v_{T_i}-u}\left(1+\frac{1}{2}w^2\rho_i^2\partial_x^2\right)e^{-u^2/2}e^{-w^2/2}dw^2}du}\right]\delta\phi \nonumber\\
    +\left[\frac{Z_ff_h}{2\sqrt{2\pi}}\frac{T_i}{T_f}\int_{-\infty}^\infty{\int_0^\infty{\frac{\omega_{*T_f,k}/k_\parallel v_{T_f}}{\omega_k/k_\parallel v_{T_f}-u}\left(1+\frac{1}{2}w^2\rho_f^2\partial_x^2\right)e^{-u^2/2}e^{-w^2/2}dw^2}du}\right]\delta\phi=0.
\end{eqnarray}
 Here, normalizations with respect to thermal speed $u\equiv v_\parallel /v_T$, $w^2\equiv 2\mu B/v_T^2$ have been used. Representing the integrals in terms of the plasma dispersion function $Z(\zeta)\equiv\frac{1}{\sqrt{\pi}}\int_{-\infty}^\infty{\frac{e^{-t^2}}{t-\zeta}dt}$, \cite{FriedBook61}
 
\begin{align}\label{eqn19}
    &\Bigg[\frac{1}{\tau}+Z_i(1-f_h)+Z_f f_h\frac{T_i}{T_f}+\frac{Z_i(1-f_h)}{2\sqrt{2}}\frac{\rho_i}{x}\frac{L_s}{L_{T_i}}\Bigl[\bigl(1-\rho_i^2\partial_x^2\bigr)Z(\zeta_i)-2\bigl(1+\rho_i^2\partial_x^2\bigr)\bigl(\zeta_i+\zeta_i^2 Z(\zeta_i)\bigr)\Bigr]\Bigg]\delta\phi \nonumber\\&\quad +\Bigg[\frac{Z_f f_h}{2\sqrt{2}}\frac{T_i}{T_f}\frac{\rho_f}{x}\frac{L_s}{L_{T_f}}\Bigl[\bigl(1-\rho_f^2\partial_x^2\bigr)Z(\zeta_f)-2\bigl(1+\rho_f^2\partial_x^2\bigr)\bigl(\zeta_f+\zeta_f^2 Z(\zeta_f)\bigr)\Bigr]\Bigg]\delta\phi=0.
\end{align}
where $\zeta_i\equiv\omega/\sqrt{2}k_\parallel v_{T_i}$ and $\zeta_f\equiv\omega/\sqrt{2}k_\parallel v_{T_f}$, and electron density profile is assumed to be nearly flat. Inverted density profile case, in which ITG instability \cite{HahmPFB89} and ensuing turbulence \cite{HahmPFB90} becomes very weak even when they are linearly unstable, requires a separate treatment.

We use the small-$\zeta$ asymptotic formula for the plasma dispersion function, $Z(\zeta)\approx i\sqrt{\pi}$. This describes the strong wave-particle resonance. Then Eq.~(\ref{eqn19}) reduces to

\begin{equation}\label{eqn20}
    \left[\frac{1}{\tau}+Z_i(1-f_h)+Z_ff_h\frac{T_i}{T_f}+i\frac{1}{2}\sqrt{\frac{\pi}{2}}Z_i(1-f_h)\frac{\rho_i}{x}\frac{L_s}{L_{T_i}}(1-\rho_i^2\partial_x^2)+i\frac{1}{2}\sqrt{\frac{\pi}{2}}Z_ff_h\frac{T_i}{T_f}\frac{\rho_f}{x}\frac{L_s}{L_{T_f}}(1-\rho_f^2\partial_x^2)\right]\delta\phi=0.
\end{equation}
This can be expressed as an eigenmode equation in Schr{\"o}dinger form
\begin{equation}\label{eqn21}
    \left[\frac{\partial^2}{\partial\tilde{x}^2}+Q(\tilde{x})\right]\delta\phi=0
\end{equation}
where a dimensionless variable $\tilde{x}\equiv x/\rho_i$ is used, and
\begin{equation}\label{eqn22}
    Q(\tilde{x})=\frac{-Z_i(1-f_h)-Z_ff_h\frac{L_{T_i}}{L_{T_f}}\sqrt{\frac{T_i}{T_f}}
    +i2\sqrt{\frac{2}{\pi}}\left[\frac{1}{\tau}+Z_i(1-f_h)+Z_ff_h\frac{T_i}{T_f}\right]\frac{L_{T_i}}{L_s}\tilde{x}}{Z_i(1-f_h)+Z_ff_h\frac{L_{T_i}}{L_{T_f}}\sqrt{\frac{T_f}{T_i}}}.
\end{equation}
This expression extends that of Ref.~\onlinecite{HahmPFB89} by including the fast ion contributions. Eigenvalue of Eq.~(\ref{eqn21}) can be obtained via WKB quantization condition,
\begin{equation}
    \int_0^{\tilde{x}_T}{\sqrt{Q(\tilde{x})}d\tilde{x}}=(2l+1)\frac{\pi}{4}
\end{equation}
where $\tilde{x}_T$ is the turning point and $l$ is the eigenmode number in radial domain. This yields the following expression,
\begin{equation}\label{eqn24}
    \frac{L_s}{L_{T_i}}=\frac{3}{2}\sqrt{\frac{\pi}{2}}\left[\frac{1}{\tau}+Z_i(1-f_h)+Z_ff_h\frac{T_i}{T_f}\right]\frac{\left(Z_i(1-f_h)+Z_ff_h\frac{L_{T_i}}{L_{T_f}}\sqrt{\frac{T_f}{T_i}}\right)^{1/2}}{\left(Z_i(1-f_h)+Z_ff_h\frac{L_{T_i}}{L_{T_f}}\sqrt{\frac{T_i}{T_f}}\right)^{3/2}}.
\end{equation}
Eq.~(\ref{eqn24}) reduces to the expression in Ref.~\onlinecite{HahmPFB89} in the absence of fast ions ($f_h\rightarrow0$). For moderately high fast ion temperature, favorable dependence of the threshold on $T_i/T_e$ has been quoted in relation to the hot ion regime in previous publications. \cite{ScottPRL90,StambaughPFB90,CasatiPoP08,ManticaPRL11,TardiniNF07,KimNF18,RenPPCF23,HeNF24}

\begin{figure}
\includegraphics[width=0.8\linewidth]{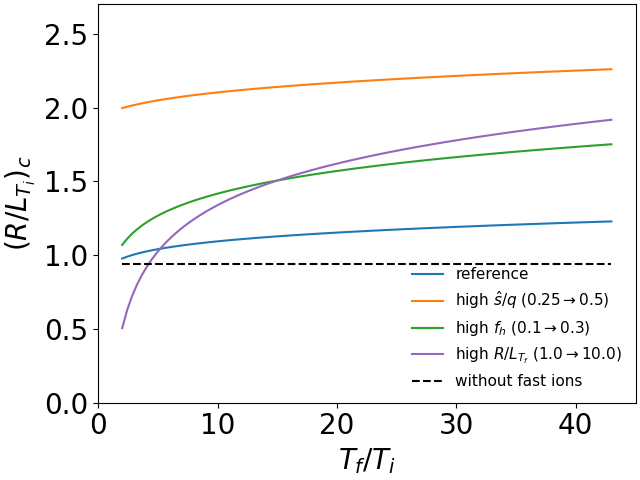}%
\caption{Linear threshold $(R/L_{T_i})_c$ in sheared slab geometry calculated by solving Eq.~(\ref{eqn24}). Reference parameters are $f_h=0.1$, $Z_i=Z_f=1$, $R/L_{T_f}=1$, $\tau=1.0$, $\hat{s}/q=0.25$, and the corresponding solution is shown in blue. Parametric dependence is checked by increasing the normalized magnetic shear to $\hat{s}/q=0.5$ (orange line), fast ion fraction to $f_h=0.3$ (green line), and normalized fast ion temperature gradient to $R/L_{T_f}=10.0$ (purple line). The linear threshold in the absence of fast ions is shown in black dashed line.\label{fig:slab}}%
\end{figure}

Fast ion effects appear in three places on the RHS of Eq.~(\ref{eqn24}). Since $L_{T_i}$ also appears on the RHS, the explicit threshold expression $\left(\frac{L_s}{L_{T_i}}\right)_c$ can be written as a solution of cubic algebraic equation. We can easily obtain its value numerically for a given set of $f_h$, $T_f/T_i$, $L_{T_f}$, etc. Here, we present a few representative results. These are shown in Fig.~\ref{fig:slab}. We note that $\left(R/L_{T_i}\right)_c$ increases gradually, but monotonically as $T_f/T_i$ increases for fixed $f_h$ and $L_{T_f}$. $Z_ff_h\frac{L_{T_i}}{L_{T_f}}\sqrt{\frac{T_f}{T_i}}$ term in the numerator on the RHS of Eq.~(\ref{eqn24}) which increases with $T_f/T_i$ is mainly responsible for this favorable behavior. However, for very high values of $T_f/T_i$, the long wavelength approximation $\left(k_\perp^2\rho_f^2\ll1\right)$ breaks down easily. Physically speaking, when $k_\perp\rho_f$ becomes large enough, fast ions' response to perturbed potential gets unmagnetized and negligible as $\Gamma_0(k_\perp\rho_f)\equiv I_0(k_\perp\rho_f)e^{-k_\perp^2\rho_f^2}\rightarrow0$, where $I_0$ is the modified Bessel function. So fast ions' effects should only appear in terms of the dilution, as their adiabatic response also becomes negligible as $T_i/T_f\rightarrow0$. Therefore, we expect the following formula without a term proportional to $\sqrt{\frac{T_f}{T_i}}$ in the numerator is a better approximation than Eq.~(\ref{eqn24}) when $k_\perp\rho_f>1$:
\begin{equation}\label{eqn25}
    \frac{L_s}{L_{T_i}}=\frac{3}{2}\sqrt{\frac{\pi}{2}}\left[\frac{1}{\tau}+Z_i(1-f_h)+Z_ff_h\frac{T_i}{T_f}\right]\frac{Z_i^{1/2}(1-f_h)^{1/2}}{\left(Z_i(1-f_h)+Z_ff_h\frac{L_{T_i}}{L_{T_f}}\sqrt{\frac{T_i}{T_f}}\right)^{3/2}}.
\end{equation}
Its asymptotic expression for $T_f/T_i\rightarrow\infty$ (for fixed $k_\perp^2\rho_i^2\ll1$) is given by 
\begin{equation}\label{eqn26}s
    \left(\frac{L_s}{L_{T_i}}\right)_c=\frac{3}{2}\sqrt{\frac{\pi}{2}}\left(1+\frac{1}{Z_i(1-f_h)\tau}\right).
\end{equation}
In this limit, only the favorable fast-ion-induced thermal ion dilution effect remains effective. This asymptotic limit is applicable to ITER plasma with 3.5 MeV fusion product $\alpha$-particles, and some of ICRH plasmas of present day large devices. \cite{MazziNP22}

\section{Strongly ballooning ion temperature gradient mode in toroidal geometry}\label{sec:toroidal}

In this section we assume $k_\parallel v_\parallel\ll\omega_{curv,k},\omega_{\nabla B,k}$.\cite{RomanelliPFB89} For feasibility of analytic progress, we make use of the constant energy resonance approximation $v_\perp^2+2v_\parallel^2=\frac{4}{3}(v_\perp^2+v_\parallel^2)$, \cite{RomanelliPFB89} thereby giving $\omega_{curv}+\omega_{\nabla B}=\frac{2}{3}\xi_{D}(v_\perp^2+v_\parallel^2)$, where $\xi_{D,k}\equiv (cm/ZeB)\bm{b}\times\nabla \ln B\cdot \bm{k}$. The expression corresponding to Eq.~(\ref{eqn16}) would be
\begin{equation}\label{eqn27}
    \delta h_k=\frac{\omega_k-\omega_{*T,k}}{\omega_k-\omega_{curv,k}-\omega_{\nabla B,k}}\frac{ZeJ_0\delta\phi_k}{T}f_0.
\end{equation}

We also employ the widely used local approximation evaluating the drift frequencies $\omega_{curv}$ and $\omega_{\nabla B}$ at $\theta=0$ (low field side mid-plane) assuming strongly ballooning mode structure. \cite{HortonPF81} Note that analytical derivations of ion temperature gradient instability threshold in toroidal geometry without consideration of fast ions have been carried out in Ref.~\onlinecite{RomanelliPFB89} and \onlinecite{YoonNF10}. Then, after a long wavelength approximation $k_\perp^2\rho_i^2\ll k_\perp^2\rho_f^2\ll1$, Eq.~(\ref{eqn15}) reduces to
\begin{eqnarray}\label{eqn28}
    \frac{1}{\tau}+Z_i(1-f_h)+Z_ff_h\frac{T_i}{T_f} \nonumber\\
    +Z_i(1-f_h)\sqrt{\frac{2}{\pi}}\int_0^\infty \frac{u^2}{u^2-\frac{3\omega_k}{2\xi_{D,i,k}v_{T_i}^2}}e^{-u^2/2}\left[\frac{3\omega_k}{2\xi_{D,i,k}v_{T_i}^2}-\frac{3}{2}\frac{R}{L_{n_i}}\left[1+\left(\frac{1}{2}u^2-\frac{3}{2}\right)\eta_i\right]\right]du \nonumber\\
    +Z_ff_h\frac{T_i}{T_f}\sqrt{\frac{m_fT_i}{m_iT_f}}\sqrt{\frac{2}{\pi}}\int_0^\infty \frac{u^2}{u^2-\frac{3\omega_k}{2\xi_{D,i,k}v_{T_i}^2}\frac{m_iZ_f}{m_fZ_i}}e^{-u^2/2\cdot m_fT_i/m_iT_f} \nonumber\\
    \left[\frac{3\omega_k}{2\xi_{D,i,k}v_{T_i}^2}\frac{Z_fT_i}{Z_iT_f}-\frac{3}{2}\frac{R}{L_{n_f}}\left[1+\left(\frac{1}{2}u^2\frac{m_fT_i}{m_iT_f}-\frac{3}{2}\right)\eta_f\right]\right]du=0.
\end{eqnarray}

By expressing the integrals in terms of the plasma dispersion function, Eq.~(\ref{eqn28}) can be written in the following form which is more compact,
\begin{eqnarray}\label{eqn29}
    \frac{1}{\tau}+Z_i(1-f_h)+Z_ff_h\frac{T_i}{T_f} \nonumber\\
    +Z_i(1-f_h)\left[-\frac{3}{2}\frac{R}{L_{T_i}}\left(\frac{1}{2}+\alpha_c+\alpha_c^{3/2}Z(\sqrt{\alpha_c})\right)+\left(2\alpha_c-\frac{3}{2}\frac{R}{L_{n_i}}+\frac{9}{4}\frac{R}{L_{T_i}}\right)\left(1+\sqrt{\alpha_c}Z(\sqrt{\alpha_c})\right)\right] \nonumber\\
    +Z_ff_h\frac{T_i}{T_f}\Biggl[-\frac{3}{2}\frac{R}{L_{T_f}}\left(\frac{1}{2}+\frac{Z_fT_i}{Z_iT_f}\alpha_c+\left(\frac{Z_fT_i}{Z_iT_f}\alpha_c\right)^{3/2}Z\left(\sqrt{\frac{Z_fT_i}{Z_iT_f}\alpha_c}\right)\right) \nonumber\\
    +\left(2\alpha_c\frac{Z_fT_i}{Z_iT_f}-\frac{3}{2}\frac{R}{L_{n_f}}+\frac{9}{4}\frac{R}{L_{T_f}}\right)\left(1+\sqrt{\frac{Z_fT_i}{Z_iT_f}\alpha_c}Z\left(\sqrt{\frac{Z_fT_i}{Z_iT_f}\alpha_c}\right)\right)\Biggr]=0.
\end{eqnarray}

\begin{figure}
\includegraphics[width=0.8\linewidth]{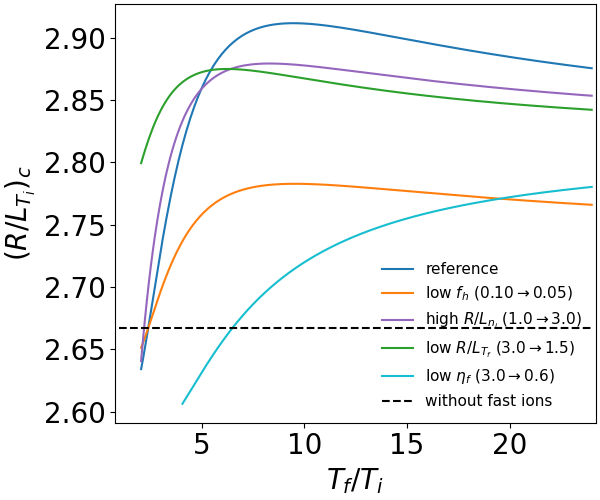}%
\caption{Linear threshold $(R/L_{T_i})_c$ of strongly ballooning mode calculated by solving Eqs.~(\ref{eqn29}) and (\ref{eqn30}). Reference parameters are $f_h=0.1$, $Z_f/Z_i=2$, $R/L_{T_f}=3.0$, $R/L_{n_f}=R/L_{n_i}=1.0$, $\tau=1.0$, and the corresponding solution is shown in blue. Parametric dependence is checked by lowering the fast ion fraction to $f_h=0.05$ (orange line), raising the normalized thermal ion density gradient to $R/L_{n_i}=3.0$ (purple line), lowering the normalized fast ion temperature gradient to $R/L_{T_f}=1.5$ (green line), and lowering the normalized fast ion temperature gradient to density gradient ratio $\eta_f$ to 0.6 (teal line). The linear threshold in the absence of fast ions is shown in black dashed line.\label{fig:toroidal}}%
\end{figure}

Requiring the imaginary part of Eq.~(\ref{eqn28}) to be zero, the Plemelj formula yields
\begin{eqnarray}\label{eqn30}
    \frac{R}{L_{T_i}}=\frac{1}{\alpha_c-\frac{3}{2}}\left(-\frac{R}{L_{n_i}}+\frac{4}{3}\alpha_c\right) \nonumber\\
    +\frac{1}{\alpha_c-\frac{3}{2}}\left(\frac{Z_f}{Z_i}\frac{T_i}{T_f}\frac{f_h}{1-f_h}\sqrt{\frac{Z_fT_i}{Z_iT_f}}e^{\alpha_c\left(1-\frac{Z_fT_i}{Z_iT_f}\right)}\left[\frac{3}{2}\left(\frac{R}{L_{T_f}}-\frac{2}{3}\frac{R}{L_{n_f}}\right)-\alpha_c\left(\frac{Z_fT_i}{Z_iT_f}\right)\left(\frac{R}{L_{T_f}}-\frac{4}{3}\right)\right]\right).
\end{eqnarray}
Here we have defined normalized resonance frequency $\alpha_c\equiv 3\omega_k/4\xi_{D,i,k}v_{T_i}^2$. Now, the linear threshold can be obtained by solving Eq.~(\ref{eqn30}) and the real part of Eq.~(\ref{eqn29}). A few representative cases are shown in Fig.~\ref{fig:toroidal}. The overall trend exhibits a local maximum of $\left(R/L_{T_i}\right)_c$ at a certain $T_f/T_i$, with the exception of low $\eta_f$ case. This optimal temperature ratio, which corresponds to the maximization of the stabilization effect of wave-particle resonance, is sensitive to the fast ion temperature gradient $R/L_{T_f}$. As the temperature ratio increases beyond this point, the threshold gradient converges toward a level set by the dilution limit (Eq.~(\ref{eqn36})). For cases with $\eta_f<2/3$, the threshold increases monotonically with the temperature ratio. The overall magnitude of the threshold gradient largely depends on fast ion density fraction $f_h$.

By substituting the resonance frequency in thermal-only case
\begin{equation}
    \alpha_c=\frac{\frac{3}{2}\frac{R}{L_{T_i}}-\frac{R}{L_{n_i}}}{\frac{R}{L_{T_i}}-\frac{4}{3}}
\end{equation}
to the thermal ion's non-adiabatic term (fourth term) of Eq.~(\ref{eqn29}), it can be simplified with reasonable accuracy to
\begin{equation}\label{eqn32}
    -\frac{3}{4}Z_i(1-f_h)\frac{R}{L_{T_i}}.
\end{equation}
For the fast ion's non-adiabatic term (fifth term), we can rearrange it to
\begin{equation}\label{eqn33}
    -\frac{3}{4}Z_ff_h\frac{T_i}{T_f}\frac{R}{L_{T_f}}+\frac{3}{2}Z_ff_h\frac{T_i}{T_f}\left[\frac{Z_fT_i}{Z_iT_f}\alpha_c\left(\frac{4}{3}-\frac{R}{L_{T_f}}\right)+\left(\frac{3}{2}\frac{R}{L_{T_f}}-\frac{R}{L_{n_f}}\right)\right]\left[1+\sqrt{\frac{Z_fT_i}{Z_iT_f}\alpha_c}Z\left(\sqrt{\frac{Z_fT_i}{Z_iT_f}\alpha_c}\right)\right]
\end{equation}
Considering the typical parameter range, we observe that
\begin{equation}\label{eqn34}
    1+\sqrt{\frac{Z_fT_i}{Z_iT_f}\alpha_c}Z\left(\sqrt{\frac{Z_fT_i}{Z_iT_f}\alpha_c}\right)=\sqrt{\frac{2}{\pi}}\int_0^\infty{\frac{t^2}{t^2-2\alpha_c\frac{Z_fT_i}{Z_iT_f}}e^{-t^2/2}dt}\sim O(0.1)
\end{equation}
This is because in the integral, there exists a partial cancellation between the negative contribution in the range $u^2<2\frac{Z_fT_i}{Z_iT_f}\alpha_c$ and the positive contribution in the range $u^2>2\frac{Z_fT_i}{Z_iT_f}\alpha_c$. For instance, in representative JET discharges with  ICRH-driven fast ions, \cite{CitrinPRL13} the evaluated numerical value of Eq.~(\ref{eqn34}) is $-0.05$ for \textsuperscript{3}He minority ICRH-heated plasma parameters ($Z_f/Z_i=2.0$, $T_f/T_i=6.9$, $\alpha_c=3.3$), and for a standard ITER scenario with ICRH-heated \textsuperscript{3}He ions, \cite{ParailNF13,DiSienaNF18} the integral evaluates to $0.08$ ($Z_f/Z_i=2.0$, $T_f/T_i=9.3$, $\alpha_c=3.4$). Therefore, this term can be ignored considering $R/L_{T_f}\sim3-20$, thus leaving only the first term of Eq.~(\ref{eqn33}). For NBI-heated D plasma parameters in JET \cite{CitrinPRL13} ($Z_f/Z_i=1.0$, $T_f/T_i=9.8$, $\alpha_c=2.0$), both the first term and the second term of Eq.~(\ref{eqn33}) turn out to be negligible compared to other terms in Eq.~(\ref{eqn29}). So keeping only the first term of Eq.~(\ref{eqn33}) does not affect the final result. We note that KSTAR FIRE mode cases \cite{HanN22,KimNF23,NaNF26} require a separate consideration due to an inverted thermal ion density profile. \cite{HahmPFB89,DuPoP17} These approximations yield a simplified expression for the resonance frequency,
\begin{equation}\label{eqn35}
    \alpha_c=\frac{3}{2}+\left(2-\frac{R}{L_{n_i}}\right)\cdot\left[\frac{4}{3}\frac{1}{Z_i(1-f_h)\tau}-\frac{Z_fT_i}{Z_iT_f}\frac{f_h}{1-f_h}\left(\frac{R}{L_{T_f}}-\frac{4}{3}\right)\right]^{-1}.
\end{equation}
Finally, a good approximation of the threshold $\left(R/L_{T_i}\right)_c$ can be routinely obtained by substituting $\alpha_c$ given by Eq.~(\ref{eqn35}) to the RHS of Eq.~(\ref{eqn30}).

As we discussed in Sec.~\ref{sec:slab}, in the limit of $T_f/T_i\rightarrow\infty$, for a fixed $k_\perp^2\rho_i^2\ll1$, fast ion response becomes unmagnetized as $k_\perp^2\rho_f^2\gg1$. Therefore only diluted the thermal ion response remains, resulting in
\begin{equation}\label{eqn36}
    \left(\frac{R}{L_{T_i}}\right)_c=\frac{4}{3}\left(1+\frac{1}{Z_i(1-f_h)\tau}\right).
\end{equation}

Having obtained compact formulas for ITG threshold in limiting cases of the flute-like mode in sheared slab geometry in Sec.~\ref{sec:slab} and strongly ballooning mode in this section, we are motivated to construct a connection formula which is valid for a wide range of parameters. Of course, while a simultaneous inclusion of both transit and drift wave-particle resonances can guarantee an accuracy, it would make further analytic progress almost impossible as such analytic work does not exist to our knowledge. Regarding this, Ref.~\onlinecite{JenkoPoP01} provides a useful guidance. Following Ref.~\onlinecite{JenkoPoP01}, we propose a connection formula as the sum of the threshold for flute-like mode in sheared slab (Eq.~(\ref{eqn26})) and that for strongly ballooning mode (Eqs.~(\ref{eqn29}) and (\ref{eqn30})). This can provide the useful advantage of facilitating one to readily map out instability threshold conditions for a broad range of parameters. Then, more accurate numerical evaluation can follow suit for specific cases of interest. In the dilution limit, we get from Eq.~(\ref{eqn26}) and Eq.~(\ref{eqn36}),
\begin{equation}\label{eqn37}
    \left(\frac{R}{L_{T_i}}\right)_c=\left(\frac{4}{3}+\frac{3}{2}\sqrt{\frac{\pi}{2}}\frac{|\hat{s}|}{q}\right)\left(1+\frac{T_i}{Z_i(1-f_h)T_e}\right).
\end{equation}

Although we cannot provide a rigorous derivation of the connection formulas, some discussion on the possible reasons behind its reasonable agreements with numerical results is appropriate here. It should be noted that if the ITG mode eigenfunction had enough flexibility to adjust its spatial structure, the actual threshold for an excitation could have been near to the minimum of the sheared-slab result and the strongly ballooning mode result, not the sum of those. Fortunately this does not happen as the eigenmode structure is constrained rather strictly. In fact, the threshold from \texttt{GENE} simulations considerably exceeds those of either the sheared slab or the strongly ballooning mode, while it agrees reasonably with the sum of those.

We emphasize that both $k_\parallel v_\parallel$ and $\omega_{D,i}$ depend not only on velocities, but also on the spatial location. $k_\parallel$ increases as a function of the distance from the rational surface, and the strong transit resonance of ITG mode occurs for $k_\parallel v_\parallel>\omega$ as demonstrated in Ref.~\onlinecite{HahmPFB89}. On the other hand, $\omega_{D,i}$ value is typically maximized at the low $B$-field side of the mid-plane ($\theta=0$). Furthermore, it can even change sign along the magnetic field, so its effect gets averaged out for flute-like modes. So the content of Sec.~{\ref{sec:slab}} and Sec.~{\ref{sec:toroidal}} covers limiting cases in which transit resonance and drift resonance effect get maximized respectively. As a consequence, the sum of respective threshold values for each limiting case seems to capture the overall parametric dependence reasonably well. Fig.~\ref{fig:ICRH} summarizes comparisons to \texttt{GENE} simulation results. \textcolor{black}{We have analyzed an evidence of wave-particle resonant interactions in velocity space where $\delta f$ from \texttt{GENE} simulation indeed exhibits significant structures along the curve (an ellipse for a given $\omega$) in ($v_\parallel,v_\perp$)-space, which correspond to drift resonance. In addition, there are visible vertical structures which are evidences of transit resonances at $v_\parallel=\omega/k_\parallel<v_{T_i}<v_{T_f}$. These indicate that two limiting cases of the wave-particle resonant interaction actually exist in \texttt{GENE} simulations. However, these are rather semi-quantitative and premature to present in this paper. Analyzing the velocity moments representing the energy exchange between particles and wave \cite{DiSienaNF18,DiSienaPoP19} in the future can provide further relevant information. Clear exhibition of the spatial dependence of transit resonance may require analyses of a global nonlinear gyrokinetic simulation since it is not obvious in the ballooning-mode-formalism-based flux tube coordinates.} We iterate that the connection formula for the threshold as a sum of the sheared slab result and the strongly ballooning mode result should be considered as an \textit{ansatz} rather than an outcome of a rigorous derivation. The ratio $k_\parallel v_\parallel/ \omega_{D,i}$ which is indicative of each term's relative importance strongly dependends on the location in phase-space. If we choose approximate values $k_\parallel\simeq 1/qR$ and $\omega_{D,i}\simeq 2k_\theta\rho_i v_{T_i}/R$ for simplicity, we obtain $k_\parallel v_\parallel /\omega_{D,i}\simeq 1/2qk_\theta\rho_i$. Therefore, transit resonance is relatively weak for high $q$, shorter wavelength, and high velocity.

We note that the effects coming from interplay between magnetic shear dependence of the geodesic curvature and the peak location of strongly ballooning mode are not captured in our model. This interplay is known to reduce the ITG growth rate for negative magnetic shear, and persists in the presence of fast ions. \cite{DiSienaPoP19} Furthermore, addressing a possible interplay between flute-like acoustic branch and strongly ballooning branch \cite{IdomuraNJP02} is beyond the scope of this paper.

\section{Numerical simulation results}\label{sec:simulation}

In this section, linear threshold gradients obtained from local linear electrostatic gyrokinetic simulations are presented. Gyrokinetic code \texttt{GENE} \cite{JenkoPoP00} is used, which solves the 5D $\delta f$ gyrokinetic Vlasov equations on a fixed grid in phase space. Flux-tube simulations with concentric circular magnetic geometry were performed with adiabatic electrons and gyrokinetic thermal and fast ions. Nominal parameters used are as follows: $B_0=1\ T$, $R_0=1.65\ m$, $m_i=m_D$, $r/R_0=10^{-5}$, $\beta=0$, $\nu_{coll}=0$, $\lambda_{Debye}=0$.

The procedure by which the threshold is calculated is as follows. First, the linear growth rate spectrum is computed at a sufficiently high $R/L_{T_i}$, and the maximum linear growth rate $\gamma_{lin,max}$ is recorded. Next, additional linear scans are performed while progressively decreasing $R/L_{T_i}$, with the corresponding maximum $\gamma_{lin,max}$ recorded at each step. Finally, a linear fit of maximum $\gamma_{lin,max}$ as a function of $R/L_{T_i}$ is performed, and the value of $R/L_{T_i}$ at which $\gamma_{lin}=0$ (the $x$-intercept) is taken as an estimate of the linear instability threshold.

\begin{figure}
\includegraphics[width=0.8\linewidth]{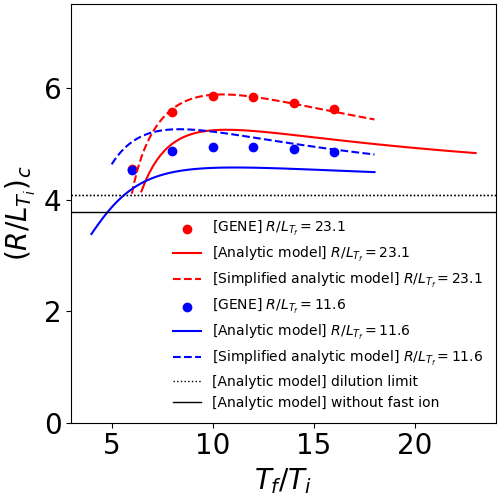}%
\caption{Comparison of the ITG linear threshold $R/L_{T_i}$ scans with two different $R/L_{T_f}$ values predicted from our analytic model (solid lines) with that obtained from GENE simulation (dots). Other parameter values are based on JET ICRH-heated discharges. Analytical threshold is calculated as the sum of the sheared-slab branch solution at dilution limit (Eq.~(\ref{eqn26})) and the strongly ballooning branch solution calculated by solving Eq.~(\ref{eqn29}) and (\ref{eqn30}). The ITG linear thresholds estimated from the simplified expression (Eq.~(\ref{eqn35})) plus the sheared-slab threshold at dilution limit are also presented (dashed lines). The threshold predicted by Eq.~(\ref{eqn37}) at dilution limit is indicated as a black dotted line, and the threshold in the absence of fast ion is shown as a black solid line. \label{fig:ICRH}}%
\end{figure}

The linear threshold dependency on $T_f/T_i$ for plasmas based on \textsuperscript{3}He minority ICRH-heated ITG-dominant JET discharge \cite{MazziNP22} is presented in Fig.~\ref{fig:ICRH}. The parameters are as follows: $f_h=0.14$, $R/L_{n_i}=1.25$, $R/L_{n_f}=1.60$, $Z_i=1$, $Z_f=2$, $\tau=1$, $q=1.7$, and $\hat{s}=0.5$. Two different $R/L_{T_f}$ values are scanned to analyze the parameter sensitivity. The linear threshold of our analytical model is calculated as the sum of the sheared-slab branch solution at dilution limit from Eq.~(\ref{eqn26}) and the solution obtained by solving Eq.~(\ref{eqn29}) and (\ref{eqn30}) with a numerical solver.

Because ICRH forms fast ions with high $\eta_f>\frac{2}{3}$, there exists a condition at which the beneficial effect of wave-particle resonance is maximized. This phenomenon is manifested as the existence of a maximum point in the $T_f/T_i$ versus $(R/L_{T_i})_c$ curve as shown in Fig.~\ref{fig:ICRH}. The typical value of the optimal temperature ratio that maximizes the threshold is found to be $\sim10$ from our parameter scan. This optimal point is accurately reproduced by our analytical model, as evidenced by the agreement between the model's prediction and GENE simulation result. \textcolor{black}{Semi-quantitative similarity of this behavior and that presented in Fig.~\ref{fig:toroidal} suggests that the underlying physics discussed in Sec.~\ref{sec:toroidal} is responsible for this trend.} As the fast ion temperature is raised past this optimal point, the linear threshold converges towards the dilution limit given by Eq.~(\ref{eqn36}). Aforementioned trends are in agreement with those of the linear growth rate reported in Refs.~\onlinecite{DiSienaNF18,DiSienaPoP19}. In addition, higher $R/L_{T_f}$, which corresponds to higher injected ICRH power, leads to increase in the linear threshold $(R/L_{T_i})_c$.

\begin{figure}
\includegraphics[width=\linewidth]{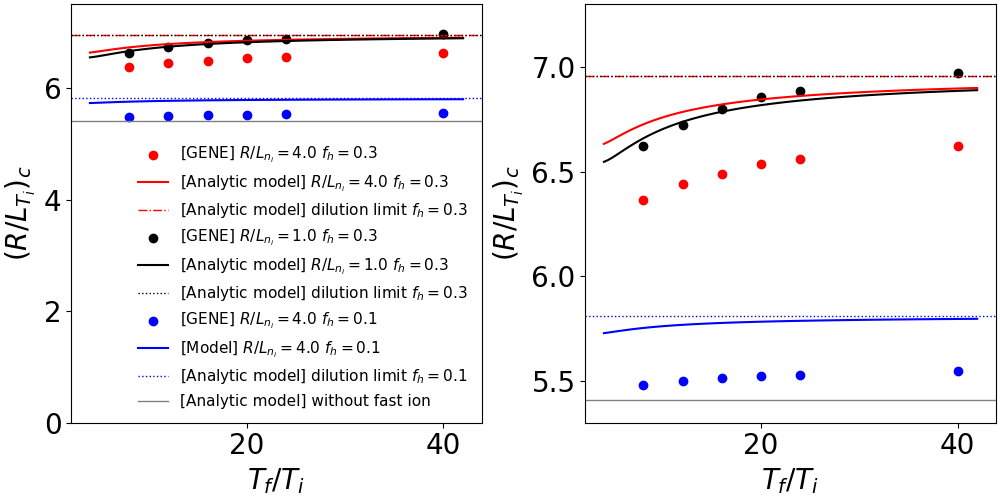}
\caption{(Left) Comparison of the ITG linear threshold $R/L_{T_i}$ scans with various $R/L_{n_i}$ and $f_h$ values predicted from our model (solid lines) with that obtained from GENE simulation (dots). Other parameters are based on ASDEX Upgrade NBI-heated discharges. Analytical threshold is calculated as the sum of the sheared-slab branch solution at dilution limit (Eq.~(\ref{eqn26})) and the strongly ballooning branch solution calculated by solving Eq.~(\ref{eqn29}) and (\ref{eqn30}). The thresholds calculated by Eq.~(\ref{eqn37}) are indicated by dashed lines, and the threshold in the absence of fast ions is shown in gray solid line. (Right) Zoomed in version of the left.}
\label{fig:NBI}
\end{figure}

Next, NBI-heated ITG-dominant ASDEX Upgrade discharge \cite{TardiniNF07} is considered. The parameters are as follows: $R/L_{n_f}=4.0$, $R/L_{T_f}=0.5$, $Z_i=Z_f=1$, $\tau=0.5$, $q=2.0$, and $\hat{s}=0.5$. $R/L_{n_i}$ and $f_h$ values are varied to observe the linear threshold's dependency on these parameters. The results are presented in Fig.~\ref{fig:NBI}. In contrast to ICRH, which achieves localized heating, NBI achieves fast ions' central fueling. Therefore, it is feasible to raise fast ions' fraction $f_h$ which results in substantial and monotonic increase of $(R/L_{T_i})_c$. Another consequence is that it can lead to a low value of $\eta_f<\frac{2}{3}$. As a result, the linear threshold shows monotonically increasing trend as $T_f/T_i$ is increased, approaching the dilution limit (Eq.~(\ref{eqn36})). From GENE results, it can be observed that near-flat thermal ion density profile ($R/L_{n_i}=1$) raises the linear threshold compared to peaked profile ($R/L_{n_i}=4$). However, this dependency is not observed in our analytic results. This is due to weak density gradient approximation used in the derivation. However, this variation is found to be rather small in relative sense, until the threshold is better quantified by $\eta_{i,crit}$ rather than $(R/L_{T_i})_c$. \cite{JenkoPoP01} In this paper, we have focused on the weak density gradient cases which are expected for fusion-relevant plasmas in future large devices. When fast ions' central fueling gets very efficient for low target density plasmas, main ion density profile can even get hollow. This has been observed in KSTAR FIRE mode plasmas. \cite{HanN22,KimNF23,NaNF26} ITG mode for the inverted density gradient has been investigated separately in the past. \cite{HahmPFB89,DuPoP17} In that case ITG turbulence is expected to be very weak even when $R/L_{T_i}$ is well above the threshold. \cite{HahmPFB90}

Based on the \texttt{GENE} results, the wavelength corresponding to the maximum linear growth rate $\gamma_{lin,max}$ is found at $k_y\rho_i\sim0.3-0.4$ for JET parameters and $k_y\rho_i\sim0.2-0.3$ for ASDEX Upgrade parameters. Assuming that $k_\perp\sim k_y$ for dominant modes, this corresponds to $k_\perp^2\rho_i^2\lesssim0.16$. Therefore, the modes contributing most significantly to the instability lie within the long wavelength regime, and the ordering $k_\perp^2\rho_i^2\ll1$ is reasonably well-satisfied for the parameters considered. This supports the consistency between the theoretical assumptions and the \texttt{GENE} simulation results. Furthermore, ITG turbulence spectrum at nonlinear saturation is well-known to peak at lower $k_\perp$ than that for the maximum linear growth rate as a consequence of the inverse spectral transfer expected for quasi-2D turbulence.

\section{Discussion and Conclusion}\label{sec:conclusion}

\begin{figure}
    \begin{tabular}{ll}
    \includegraphics[width=0.5\linewidth]{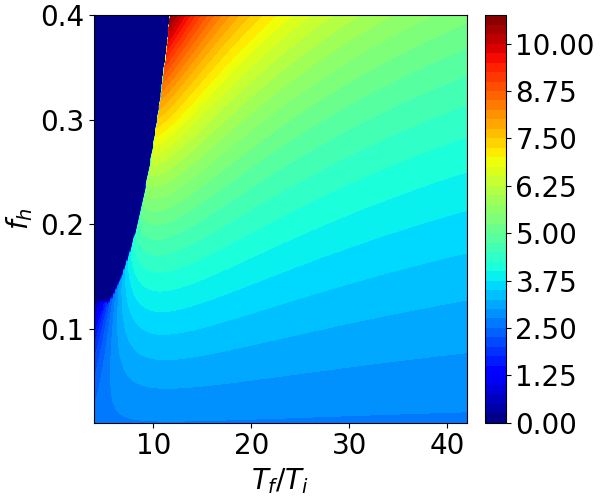}
    &
    \includegraphics[width=0.5\linewidth]{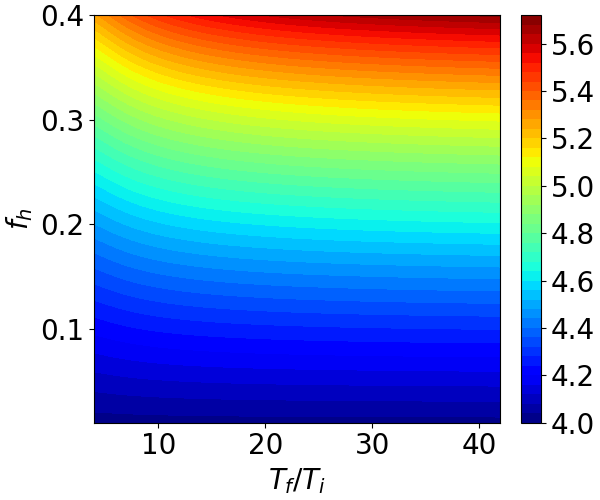}
    \end{tabular}
    \centering
    \caption{Two dimensional parameter scan of linear threshold $(R/L_{T_i})_c$ as a function of the temperature ratio $T_f/T_i$ and fast ion fraction $f_h$, obtained by solving Eqs.~(\ref{eqn29})-(\ref{eqn30}). (Left) Results from scans based on JET ICRH-heated plasma parameters, characterized by $\eta_f>\frac{2}{3}$. Other parameter values are identical to \textsuperscript{3}He minority ICRH-heated JET discharge described in Sec.~\ref{sec:simulation}. The region in the upper left of the figure corresponds to $(R/L_{T_i})_c=0$ since ITG is unstable regardless of $R/L_{T_i}$ due to high $R/L_{T_f}$. (Right) Results from scans based on ASDEX Upgrade NBI-heated plasma parameters with $\eta_f<\frac{2}{3}$. All parameter values can be found in Sec.~\ref{sec:simulation}.}
    \label{fig:overview}
\end{figure}

In this paper, we have investigated the fast ion effects on the threshold conditions of ITG mode both analytically and numerically using the gyrokinetic equation. Obviously, the threshold conditions are largely determined by the kinetic wave-particle resonance effects. The overall picture we get from Fig.~\ref{fig:overview} summarizing the behavior of $(R/L_{T_i})_c$ as a function of $f_h$ and $T_f/T_i$ are as follows:
\begin{enumerate}
    \item[(i)] The linear threshold scales favorably with the fast ion density fraction $f_h$. Increasing $f_h$ leads to substantial raising of $(R/L_{T_i})_c$ when other parameters are kept constant, as evident from greater variation of the threshold in the $y$-direction in both plots in Fig.~\ref{fig:overview}.
    \item[(ii)] Effect of $T_f/T_i$ can be classified into two branches depending on whether $\eta_f=L_{n_f}/L_{T_f}$ exceeds a critical value $\frac{2}{3}$, as shown by the difference between the two plots in Fig.~\ref{fig:overview}. If $\eta_f>\frac{2}{3}$, $(R/L_{T_i})_c$ increases for low to moderate values of $T_f/T_i$, peaks at $T_f/T_i\sim 10$, and converges towards the dilution limit (Eq.~(\ref{eqn37})) as $T_f/T_i$ is further increased. On the other hand, if $\eta_f<\frac{2}{3}$, $(R/L_{T_i})_c$ shows monotonic trend as it approaches the dilution limit at high $T_f/T_i$. Considering that $\eta_f$ is largely determined by the fast ion source (e.g. ICRH and NBI), this result has notable implications for tokamak experiments.
\end{enumerate}

In addition, $(R/L_{T_i})_c$ and $\hat{s}/q$ are positively correlated, meaning that the linear threshold is raised when the normalized magnetic shear is increased (see Fig.~\ref{fig:slab}). For $T_f/T_i\gg1$, finite Larmor radius effects for thermal ions and fast ions are manifested in opposite asymptotic limits, (i.e., $k_\perp\rho_i\ll1\ll k_\perp\rho_f$). In particular, fast ion response gets unmagnetized and negligible. In this regime, fast ions' contribution is restricted to the dilution of thermal ion species (see Figs.~\ref{fig:slab}-\ref{fig:NBI}). This asymptotic limit is relevant to high energy fusion products such as 3.5 MeV $\alpha$-particles in ITER and future devices.

For further improvement of the accuracy of our analytic results, a connection formula as a weighted sum of sheared slab threshold and that in the strongly ballooning limit with coefficients reflecting the validity regime of transit frequency-dominant and drift frequency-dominant resonance respectively could be constructed. \textcolor{black}{For this, a velocity coordinate transform which simplifies the mathematical expression for a combination of the transit and drift resonances \cite{KimPoP94} might help.} However, that is beyond the scope of this paper and can possibly be pursued in the future.

As mentioned in the introduction, the subject covered in this paper is one of the conceptually simplest considerations addressing the fast ion effects on turbulence. Various nonlinear physics \cite{NaNRP25} involving self-generated structures such as zonal flows as influenced by fast ions \cite{HahmPoP23,ChoiNF24} could play prominent roles in many cases. Nonetheless, the linear results presented in this paper can still have a significant impact, being useful for fair assessment of important physics in various cases.

In addition, electromagnetic effects associated with finite plasma $\beta$, which can increase with fast ion fraction, are neglected in the present electrostatic model. Such effects are generally expected to provide additional stabilization of ITG modes \cite{CitrinPPCF15} and may therefore quantitatively modify but not qualitatively alter the favorable trends reported here. Furthermore, the present analysis assumes adiabatic electrons, while kinetic electron dynamics including trapped electron effects can modify ITG stability and introduce additional modes such as trapped electron mode (TEM). Therefore, the trends identified here are expected to remain applicable only in ITG-dominant regimes, such as the JET \cite{MazziNP22} and ASDEX Upgrade \cite{TardiniNF07} discharges considered in this paper. For TEM-dominant regimes which can occur for electron-heated plasmas, a separate study is required which we don't pursue in this paper.

Finally, we expect that the beneficial effects of fast ions on ITG stability are likely to dominate over the unfavorable effects on ETG stability reported in Appendix. With its natural electron gyroradius length scale for linear modes, ETG's contribution to turbulent transport is expected to be much smaller than that from unstable ITG or TEM. ETG-driven transport would be considerable only when the radially elongated streamers are generated. However, such structures can be destroyed not only by zonal flow shear, \cite{DiamondPPCF05} but also by the vortex flow shear associated with larger structures. Examples of the latter include the ITG mode, \cite{HollandPRA05} large magnetic island, \cite{IshizawaPPCF19,HahmPoP21,ChoiPRL22,YoonNF24} and toroidal Alfv{\'e}nic eigenmode (TAE). \cite{IshizawaCP25} On the other hand, ETG mode is also known to influence larger scale turbulence and overall outcome of cross-scale interactions of turbulence can be beneficial. \cite{MaeyamaNC22}

\newpage

\begin{acknowledgments}
We acknowledge useful discussions with S. J. Park, A. Ishizawa, and H.-T. Kim. This work was supported by the National Research Foundation of Korea (NRF) grant funded by the Korea Government (MSIT) (Grant No. 2023R1A2C1007735\textcolor{black}{, Grant No. RS-2026-25484551} and Grant No. RS-2024-00358933). The simulations presented in this work were performed using the HPC resources from KFE KAIROS. The authors also thank the Research Institute of Energy and Resources and the Institute of Engineering Research at Seoul National University.
\end{acknowledgments}

\appendix*

\section{Analytical linear threshold condition for electron temperature gradient mode in the presence of fast ions}\label{sec:ETG}

Since its radial width expected from the linear theory scales with the electron gyroradius, anomalous electron thermal transport caused by ETG mode is not significant unless it develops radially elongated streamers nonlinearly. \cite{JenkoPoP00} This topic has been recently reviewed \cite{RenRMPP24} including progress in experimental measurements. \cite{MazzucatoPRL08}
It is known that linear properties of ETG mode in the limit of adiabatic ion response are isomorphic to those of ITG (ion temperature gradient) mode, with the roles of ions and electrons reversed. Ref.~\onlinecite{JenkoPoP01} carried out \texttt{GENE} gyrokinetic simulation scan of the instability threshold, and found that the linear combination of analytical formulas from Ref.~\onlinecite{HahmPFB89} and Ref.~\onlinecite{RomanelliPFB89} provides a good agreement for weak density gradient.
\begin{equation}\label{eqnA1}
    \left(\frac{R}{L_{T_e}}\right)_c=\left(\frac{4}{3}+\frac{3}{2}\sqrt{\frac{\pi}{2}}\frac{\hat{s}}{q}\right)\left(1+\frac{T_e}{T_i}\right)
\end{equation}

Here, we assume the responses of thermal and fast ions are strongly unmagnetized with $k_\perp^2\rho_f^2\gg k_\perp^2\rho_i^2\gg1$, and therefore, adiabatic. Then the ETG-equivalent of Eq.~(\ref{eqn15}) would be
\begin{equation}
    \left[1+Z_i(1-f_h)\frac{T_e}{T_i}+Z_ff_h\frac{T_e}{T_f}\right]\frac{n_{e0}e\delta\phi}{T_e}-2\pi\int{B^*\delta h_e J_0d\mu dv_\parallel}=0.
\end{equation}
Following the same procedure shown in Sec.~\ref{sec:slab}, the eigenmode equation becomes
\begin{equation}
    \left[\frac{\partial^2}{\partial x^2}+Q(x)\right]\delta\phi=0
\end{equation}
where
\begin{equation}
    Q(x)=-1+i2\sqrt{\frac{2}{\pi}}\left[1+Z_i(1-f_h)\frac{T_e}{T_i}+Z_ff_h\frac{T_e}{T_f}\right]\frac{L_{T_e}}{L_s}x.
\end{equation}
Solving for the eigenvalue via the WKB quantization condition yields
\begin{equation}\label{eqnA5}
    \frac{L_s}{L_{T_e}}=\frac{3}{2}\sqrt{\frac{\pi}{2}}\left[1+Z_i(1-f_h)\frac{T_e}{T_i}+Z_ff_h\frac{T_e}{T_f}\right].
\end{equation}

On the other hand, for the toroidal branch, Eq.~(\ref{eqn28}) becomes
\begin{equation}\label{eqnA6}
    1+Z_i(1-f_h)\frac{T_e}{T_i}+Z_ff_h\frac{T_e}{T_f}+\sqrt{\frac{2}{\pi}}\int_0^\infty \frac{u^2}{u^2-2\alpha_c}e^{-u^2/2}\left[2\alpha_c-\frac{3}{2}\frac{R}{L_{n_e}}\left[1+\left(\frac{1}{2}u^2-\frac{3}{2}\right)\eta_e\right]\right]du=0
\end{equation}
where $\alpha_c\equiv 3\omega_k/4\xi_{D,e,k}v_{T_e}^2$. Utilizing Plemelj formula gives
\begin{equation}\label{eqnA7}
    \alpha_c=\frac{\frac{3}{2}\frac{R}{L_{T_e}}-\frac{R}{L_{n_e}}}{\frac{R}{L_{T_e}}-\frac{4}{3}},
\end{equation}
and substitution into Eq.~(\ref{eqnA6}) yields the threshold condition:
\begin{equation}\label{eqnA8}
    \frac{R}{L_{T_e}}=\frac{4}{3}\left(1+Z_i(1-f_h)\frac{T_e}{T_i}+Z_ff_h\frac{T_e}{T_f}\right).
\end{equation}

\begin{figure}
\includegraphics[width=0.8\linewidth]{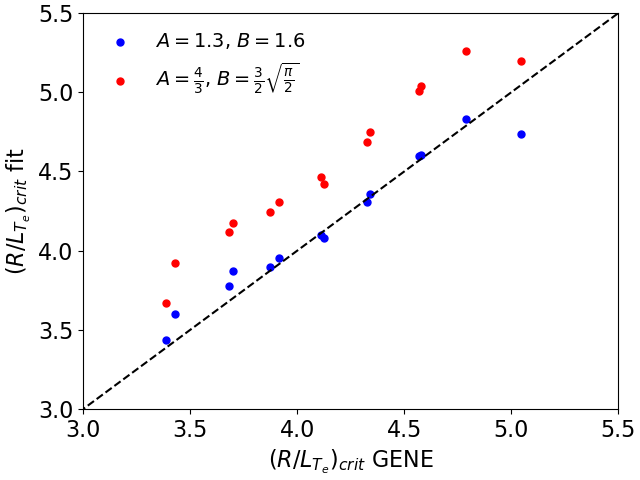}%
\caption{Comparison of linear threshold $R/L_{T_e}$ of ETG instability obtained from GENE simulation with the analytic linear threshold from Eq.~(\ref{eqnA9}). Blue dots compares GENE results with the line of best fit, which corresponds to the coefficient set $A=1.3$, $B=1.6$. Red dots compares GENE results with the coefficients $A=\frac{4}{3}$, $B=\frac{3}{2}\sqrt{\frac{\pi}{2}}$ obtained from analytic derivation. \label{fig:ETG}}%
\end{figure}

Finally, we propose the linear threshold of ETG, as the linear combination of Eq.~(\ref{eqnA5}) and (\ref{eqnA8}):
\begin{equation}\label{eqnA9}
    \left(\frac{R}{L_{T_e}}\right)_c=\left(A+B\frac{|\hat{s}|}{q}\right)\left(1+Z_i(1-f_h)\frac{T_e}{T_i}+Z_ff_h\frac{T_e}{T_f}\right)
\end{equation}
It is noteworthy that both high fast ion temperature and dilution destabilize ETG. This trend is similar to previous publications addressing the behavior of $\gamma_{lin}$. \cite{BonanomiNF18,KimNF24} Gyrokinetic simulation scan with GENE code confirmed Eq.~(\ref{eqnA9}) as shown in Fig.~\ref{fig:ETG}, with the coefficient set $A=1.3$, $B=1.6$ yielding the best fit, while the analytic derivation yields $A=\frac{4}{3}$ and $B=\frac{3}{2}\sqrt{\frac{\pi}{2}}$. Coefficient discrepancies from Eq.~(\ref{eqnA1}) can be understood from different geometries employed as explained in Ref.~\onlinecite{LapillonnePoP09}.

\bibliography{reference}

\clearpage

\end{document}